\documentclass[times]{akari2017} 

\shorttitle{A new parameter in the FMR of star-forming galaxies}
\shortauthors{Hashimoto et al.}

\begin{document}

\title{Surface density: a new parameter in the fundamental metallicity relation of star-forming galaxies} 



\correspondingauthor{Tetsuya Hashimoto}
\email{tetsuya@phys.nthu.edu.tw}

\author{Tetsuya Hashimoto} 
\affiliation{National Tsing Hua University} 
\author{Tomotsugu Goto} 
\affiliation{National Tsing Hua University} 
\author{Rieko Momose} 
\affiliation{National Tsing Hua University} 



\begin{abstract}
Star-forming galaxies display a close relation among stellar mass, metallicity and star-formation rate (or molecular-gas mass).
This is known as the fundamental metallicity relation (FMR) (or molecular-gas FMR), and it has a profound implication on models of galaxy evolution. 
However, there still remains a significant residual scatter around the FMR. 
We show here that a fourth parameter, the surface density of stellar mass, reduces the dispersion around the molecular-gas FMR. 
In a principal component analysis of 29 physical parameters of 41,338 star-forming galaxies, the surface density of stellar mass is found to be the fourth most important parameter. 
The new four-dimensional (4D) fundamental relation forms a tighter hypersurface that reduces the metallicity dispersion to 50\% of that of the molecular-gas FMR. 
We suggest that future analyses and models of galaxy evolution should consider the FMR in a 4D space that includes surface density. 
The dilution time scale of gas inflow and the star-formation efficiency could explain the observational dependence on surface density of stellar mass.
AKARI is expected to play an important role in shedding light on the infrared properties of the new 4D FMR.
\end{abstract}



\setcounter{page}{1}



\section{Introduction} \label{Introduction} 
The relation between stellar mass and gas-phase metallicity, which was discovered a decade ago, is one of the most fundamental relations for star-forming galaxies both in the local universe and at high redshifts \citep{Tremonti2004, Erb2006}.
The gas-phase metallicity, here, is a measure of the relative amount of oxygen and hydrogen present.
Many explanations for this relation have been proposed, including metal depletion by enriched gas outflow, which depends on the galactic gravitational well \citep{Garnett2002,Tremonti2004,Koppen2007,Brooks2007,Dalcanton2007,Calura2009,Finlator2008}. 
Since its discovery, this relation has been developed into what is known as the fundamental metallicity relation (FMR) among stellar mass, metallicity and star-formation rate (SFR), defining a curved surface in the equivalent three-dimensional (3D) parameter space \citep{Mannucci2010}. 
Recent instances of the FMR based on large statistical single-fiber samples of star-forming galaxies have dramatically advanced our understanding of galaxy evolution.
For example, the negative correlation that exists between SFR and metallicity in the FMR can be understood in the context of the inflow of metal-poor gas that directly dilutes the gas-phase metallicity and also triggers star formation \citep{Mannucci2010,Forbes2014}. 

There remains a significant residual scatter in metallicity beyond measurement error even after the SFR dependence is removed \citep{Salim2014,Salim2015}.
This suggests that there could be an other factor which regulates metallicity of star-forming galaxies except for SFR.
The goal of this paper is finding out the fourth most important physical parameter of star-forming galaxies.

\section{Sample selection}  \label{Sample selection}
The galaxies in our sample were selected from the Sloan Digital Sky Survey (SDSS) Data Release 7 \citep{Abazajian2009}.
We used the publicly available catalogue of emission-line fluxes produced in collaboration between the Max Planck Institute for Astrophysics (MPA) and Johns Hopkins University (JHU) \citep{Kauffmann2003,Brinchmann2004,Salim2007}.
The criteria for selecting star-forming galaxies were based on previous research on the stellar-mass-metallicity relation \citep{Kewley2008}.
A signal-to-noise ratio (S/N) of at least 8 was required in the strong optical emission lines of [O~{\sc ii}]$\lambda$3726, 3729, H$\beta$, [O~{\sc iii}]$\lambda$5007, H$\alpha$, [N~{\sc ii}]$\lambda$6584 and [S~{\sc ii}]$\lambda$6717, 6731 for reliable estimates of metallicity and electron density.
This S/N limit on emission lines ensures accurate measurements of physical parameters but could lead to a biased sample.
The redshift range was limited to 0.04-0.1.
The lower redshift limit ensured at least 20\% higher fibre covering fraction of the total photometric $g'$-band light as required for metallicity to approximate the global value \citep{Kewley2005}.
The upper redshift limit is due to the fact that the SDSS sample of star-forming galaxies becomes incomplete at redshifts above 0.1 \citep{Kewley2006}.
Host galaxies of active galactic nuclei (AGNs) were excluded by using the emission-line-ratio diagnostic diagram (i.e. a Baldwin-Phillips-Terlevich diagram \citep{Baldwin1981}) and the AGN selection criteria \citep{Kauffmann2003}.
The resulting sample contained 41,338 star-forming galaxies after removing duplicates in the emission-line catalogue.

In total, 18 physical parameters of the sampled galaxies were compiled from various literatures, and a further 11 were calculated in this work.
Note that 11 parameters include N2O3 index which is defined as N2O3 $=\log$[([N~{\sc ii}]$\lambda 6584/{\rm H}\alpha$)/([O~{\sc iii}]$\lambda 5007/{\rm H}\beta$)].
This emission-line ratio is linked to metallicity according to the calibration formula of 12+log(O/H)=8.73+0.32 $\times$ N2O3 with a statistical uncertainty of $\sim$ 0.1 dex \citep{Pettini2004, Marino2013}.
Throughout this paper, we use N2O3 as an indicator of metallicity to avoid this large statistical uncertainty which is involved when N2O3 is converted to the actual value of metallicity.

\begin{table*}
	\centering
	\caption{
    Summary of 29 parameters of SDSS star-forming galaxies, categorised into six groups in order of radial distance in the 3D space of the factor loadings of PC1, PC2 and PC3.
    }
	\label{tab1}
      \begin{tabular}{|ll|ll|ll|}\hline
\multicolumn{2}{|c|}{MASS} & \multicolumn{2}{|c|}{METAL} & \multicolumn{2}{|c|}{ACTIVITY} \\ \hline
$M_{*}$& Stellar mass & N2O3& Metallicity indicator & $M_{\rm H_{2}}$& Molecular-gass mass  \\
M$_{i}$& $i$-band absolute magnitude & $M_{metal}$& Metal mass in ionised gas & SFR& Star-formation rate \\
M$_{z}$& $z$-band absolute magnitude & Av& Dust extinction & M$_{u}$& $u$-band absolute magnitude \\
M$_{r}$& $r$-band absolute magnitude & & & sSFR& Specific star-formation rate \\
M$_{g}$& $g$-band absolute magnitude & & & $M_{\rm HI}$& Atomic hydrogen gas mass  \\
$M_{virial}$& Virial mass ($2r_{half}\sigma^{2}/G$) & & & g-r& Colour \\
$\sigma$& Velocity dispersion & & & D4000& Strength of 4000-\AA\ break \\
$M_{igas}$& Ionised-gas mass & & & EW$_{\rm H\alpha}$& Equivalent width of H$\alpha$ \\\hline \hline
\multicolumn{2}{|c|}{SIZE/MORPHOLOGY} & \multicolumn{2}{|c|}{ENVIRONMENT} & \multicolumn{2}{|c|}{OTHER} \\ \hline
$\Sigma_{M_{*}}$& Surface density of M$_{*}$ & $M_{halo}$& Dark-matter halo mass & q& [O~{\sc iii}]$\lambda$5007/[O~{\sc ii}]$\lambda$3727 \\
$r_{half}$& $r$-band half-light radius & $\delta_{5}$& Local galaxy number density & $z$& Redshift \\
$\Sigma_{\rm SFR}$& Surface density of SFR & & & $n_{e}$& Electron density \\
$r_{disk}$& $r$-band disk radius & & & & \\
B/T& Bulge-to-total fraction & & & & \\ \hline
  \end{tabular}
\end{table*}

\section{Tightnesses of FMRs defined by various parameters}  \label{Tightnesses of FMRs}
To find a new fundamental parameter for the FMR, we carried out principal component analysis (PCA) on 41,338 star-forming galaxies selected from the SDSS Data Release 7 \citep{Abazajian2009}. 
We included 29 physical parameters categorised into six groups, namely, MASS, METAL, ACTIVITY, SIZE/MORPHOLOGY, ENVIRONMENT and OTHER (see Table \ref{tab1}.
The resulting \lq factor loadings\rq are indicators of how strongly each original parameter contributes to the new axes.
We adopt a criteria to select important parameters based on the distance in the 3D space of factor loadings of PC1, PC2 and PC3 (instead of using values along each PCA axis).
The PC1, PC2 and PC3 axes explain most ($\sim$70\%) of the overall distribution of galaxies, suggesting the contribution from the fourth parameter to these three axes if it exists.
Therefore the comprehensive contributions of the original parameters to the galactic distribution are indicated by the radial distances from the origin in the 3D space of the PC1, PC2 and PC3 factor loadings.
The physical parameters in each categories are listed in Table \ref{tab1} in order of the distance in PC1-3 factor loading space.
The most important parameters in each group (i.e. the most distant parameters in the 3D factor-loading space) are stellar mass (MASS), N2O3 (METAL), molecular-gas mass (ACTIVITY), surface density of stellar mass (SIZE/MORPHOLOGY), dark-matter halo mass (ENVIRONMENT) and ionisation parameter (OTHERS).
We conduct further exploration of these parameters as well as SFR and half-light radius, the latter two of which are included as existing references. 

\begin{figure}[ht]
\begin{minipage}[b]{0.5\linewidth}
   \centering
   \resizebox{0.8\hsize}{!}{
      \includegraphics*{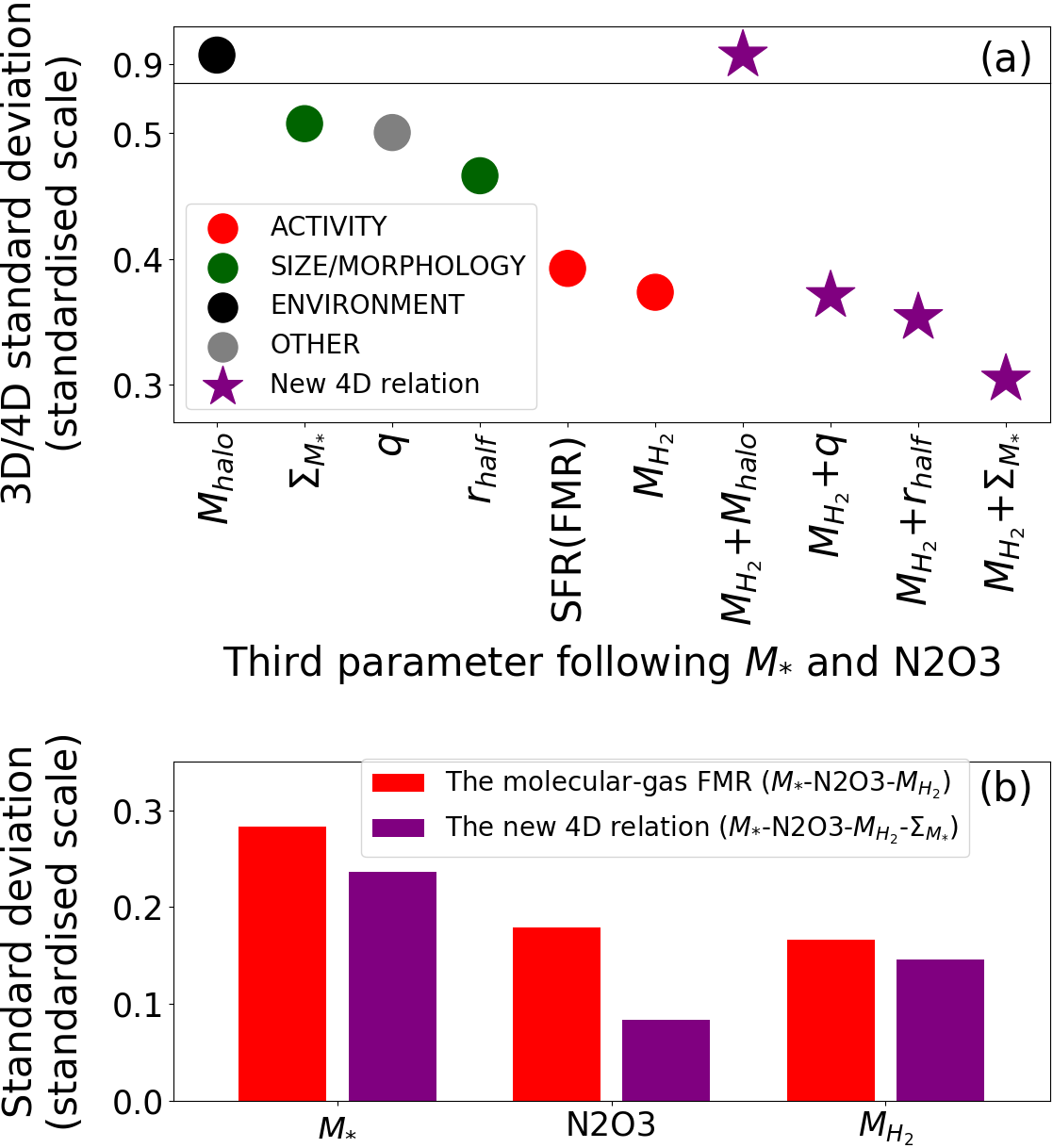}
   }
   \caption{
Standard deviations around fundamental relations in the standardised scale.
    \textbf{(a),} Standard deviations around fundamental relations in 3D parameter spaces defined by stellar mass, N2O3 and various third parameters (red, green, black and grey dots), also including 4D parameter spaces (purple stars).
    \textbf{(b),} Breakdown of standard deviations around the 3D and 4D relations (i.e. the molecular-gas FMR and the addition of the surface density of stellar mass) into triaxial components of stellar mass, N2O3 and molecular-gas mass.
   }\label{fig2}
\end{minipage}
\hspace{0.5cm}
\begin{minipage}[b]{0.5\linewidth}
   \centering
   \resizebox{0.8\hsize}{!}{
       \includegraphics*{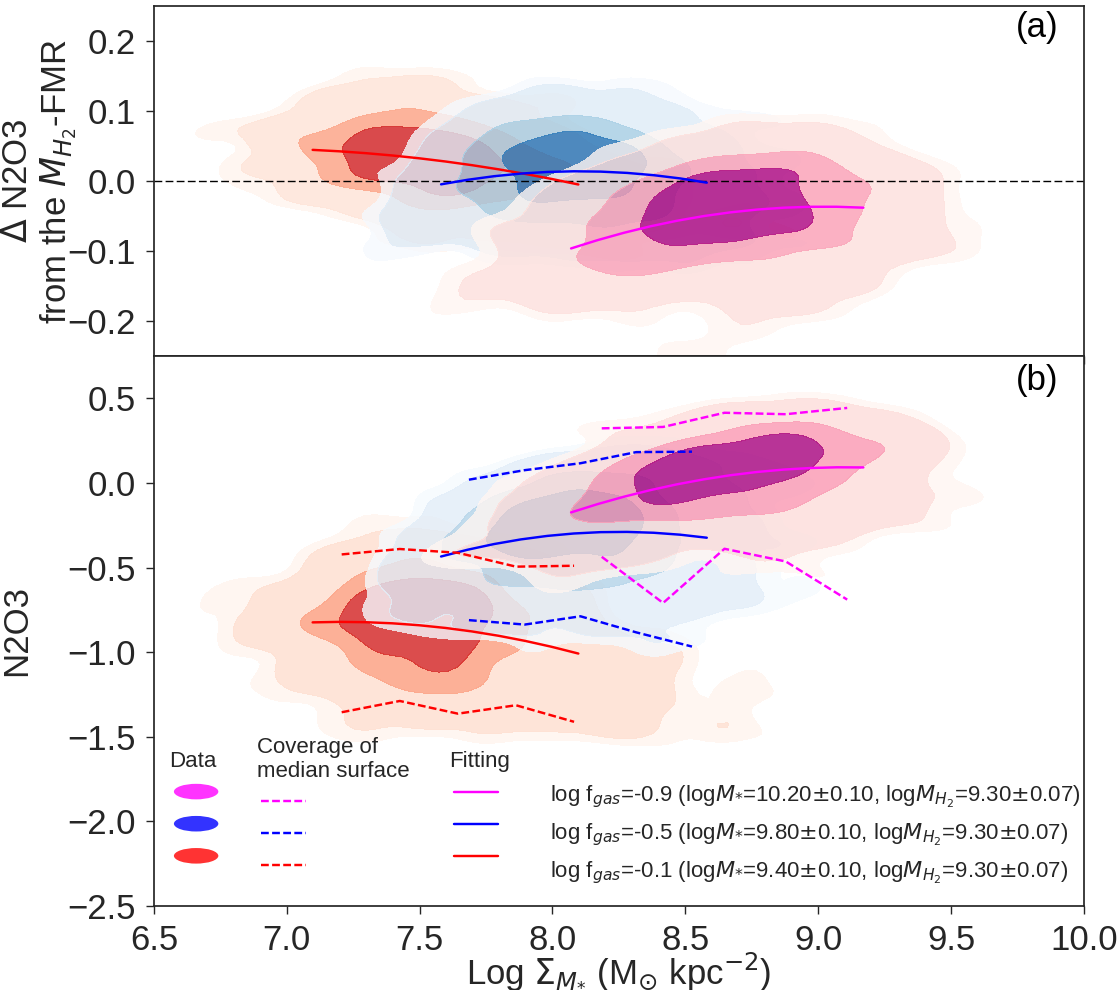}
   }
   \caption{
N2O3 dependency on the surface density of stellar mass in the physical scale.
    \textbf{(a),} N2O3 offset around the molecular-gas FMR, i.e. the 3D relation of stellar mass, N2O3 and molecular-gas mass as a function of surface density of stellar mass.
    The observed data are sampled from gas-mass-fraction bins.
    Galaxies with three different gas mass fractions are displayed in magenta, blue and red.
    Contours contain 38.2\%, 68.3\%, 95.5\% and 99.7\% of each binned sample.
    Solid lines are polynomial fits of each binned sample.
    Horizontal dashed line corresponds to the surface of the molecular-gas FMR defined by the median distribution of data points in the 3D space.
    \textbf{(b),} Same as (a) but for N2O3 in the vertical axis.
    The two dashed lines of each binned sample indicate the coverage of the projected 4D fundamental relation defined by median data points within the binning size of the sampled data.
    The lower and upper dashed lines are 1 and 99 percentiles of the N2O3 of the sampled galaxies expected from the 4D fundamental relation.
   }\label{fig3}
\end{minipage}
\end{figure}

To derive the 3D fundamental relations defined by various combinations of candidate parameters, we calculated the median surface distributions of data points.
Figure \ref{fig2}a shows the standard deviations around the median surfaces of the fundamental galactic relations defined by stellar mass, N2O3 and various choices for the third parameter.
The standard deviations were calculated using orthogonal distances from the median surfaces in the standardised 3D parameter spaces so that the tightness of the different relations could be fairly compared with each other.
A set of more fundamental parameters should construct a tighter relation according to the underlying physical mechanism of star-forming galaxies.
The use of SFR as the third parameter corresponds to the conventional FMR and shows the second smallest scatter among the various choices for the third parameter (coloured dots in Figure \ref{fig2}a).
In Figure \ref{fig2}a, the molecular-gas FMR actually results in slightly less scatter than does the conventional FMR, which may support the use of molecular-gas mass as the third fundamental parameter. 

The standard deviation of the 3D relation defined by stellar mass, N2O3 and surface density of stellar mass (left green dot in Figure \ref{fig2}a) is not as small as that by using molecular-gas mass, confirming that molecular-gas mass is the third parameter. 
However, if surface density of stellar mass is incorporated into the molecular-gas FMR as an additional fourth parameter, the extent of galactic dispersion is indeed reduced (rightmost purple star in Figure \ref{fig2}a). 
The standard deviations around the molecular-gas FMR and this new 4D relation including surface density of stellar mass are separated into the triaxial components of stellar mass, N2O3 and molecular-gas mass in the standardised scale in Figure \ref{fig2}b.
The reduction in dispersion by adding the fourth parameter is predominant in N2O3; the reductions in the other parameters are more moderate. 

We suggest the correlation between surface density of stellar mass and N2O3 as the primary cause of the residual scatter around the molecular-gas FMR. 
This is confirmed clearly in Figure \ref{fig3}a, where we show the projected N2O3 offsets from the molecular-gas FMR as a function of surface density of stellar mass.
Here the molecular-gas FMR means the median surface of the data distribution in the 3D space of stellar-mass, N2O3 and molecular-gas mass.
The vertical axis is the orthogonal offset from the molecular-gas FMR projected on N2O3 axis.

In Figure \ref{fig3}a, galaxies are sampled into three different stellar-mass bins and a fixed molecular-gas mass (i.e. different gas mass fractions $f_{gas}=M_{H_{2}}/M_{*}$) to minimise the effects of stellar mass and molecular-gas mass on N2O3.
High-$f_{gas}$ galaxies show a negative correlation between the surface density of stellar mass and N2O3 offset, whereas low-$f_{gas}$ galaxies show a positive relation within the N2O3 dispersion around the molecular-gas FMR. 
The vertical dynamic ranges of these correlation/anti-correlation are larger than the typical observational error of N2O3 in our sample ($\sim$0.015 dex derived from emission-line fluxes and their errors).

N2O3 (not offset) as a function of surface density of stellar mass is shown in Figure \ref{fig3}b for the three galaxy samples.
According to the molecular-gas FMR, the N2O3 of the samples should not show any surface-density dependence because of the fixed stellar mass and molecular-gas mass.

However, in Figure \ref{fig3}b, the three samples display the same trend, namely, a negative relation for high-$f_{gas}$ galaxies and a positive relation of low-$f_{gas}$ galaxies, which is consistent with Figure \ref{fig3}a.

\section{Physical interpretations}  \label{Physical interpretations}
If the dilution time scale is longer than the dynamical time scale of infall (i.e. $t_{dil} > t_{dyn}$), the galaxy is in the process of returning to the equilibrium metallicity as determined by the dilution and chemical enrichment by newly invoked star formation.
Such a galaxy might have observationally lower metallicity compared with the equilibrium metallicity because the bulk of emission lines is emitted from the star-forming regions in metal-poor infall gas. 
Because the ratio $t_{dil}/t_{dyn}$ is proportional to $\sqrt \rho$, where $\rho$ is the mass density \citep{Ellison2008}, galaxies with higher surface mass density may be expected to have larger values of $t_{dil}/t_{dyn}$, suggesting lower metallicity.
This time lag before metallicity dilution is expected to be more evident in gas-rich galaxies that are undergoing active infall and star formation.
This is reflected exactly in the negative correlation for high-$f_{gas}$ galaxies shown in Figure \ref{fig3}, whereas the low-$f_{gas}$ galaxies indicate the opposite trend of this metallicity-dilution scenario.
The latter type of galaxy is probably in the terminal stage of molecular-gas depletion, which means that its metallicity is already in equilibrium.

Star-formation efficiency is a likely reasonable explanation for the surface-density-N2O3 relation of low-$f_{gas}$ galaxies.
The Schmidt-Kennicutt law indicates that a more concentrated galaxy has a higher SFR compared with the molecular-gas mass, in other words a higher SFR/$M_{H_{2}}$ ratio.
This means that galaxies with higher surface density are more efficient at turning gas into stars, resulting in higher metallicity.
This expected trend is consistent with the observed surface-density-N2O3 relation of low-$f_{gas}$ galaxies shown in Figure \ref{fig3}.
Star-formation efficiency has been discussed as an alternative to the aforementioned galactic-wind scenario as a means to reproduce the stellar-mass-metallicity relation \citep{Brooks2007,Dalcanton2007,Calura2009}.
Massive galaxies form stars more effectively and are more chemically evolved than do less massive galaxies, which is known as the \lq downsizing\rq formation of galaxies. 
Such downsizing is related to the total stellar mass of the galaxy.
The surface-density-N2O3 relation suggests that galaxies with higher surface density form stars more effectively and are more chemically evolved than do galaxies with lower surface density for a fixed stellar mass. 
The present work is the first time that the relative dependence of the full FMR is shown quantitatively.
Many physical parameters in our analysis are based on optical properties of star-forming galaxies.
Near infrared properties of the new 4D FMR could reveal detailed physics of dust and obscured star formation.
AKARI is expected to play an important role in shedding light on the infrared properties of the new 4D FMR.





\end{document}